\documentclass[twocolumn,10pt,prl]{revtex4}

\def\mysections#1{{\bf #1.} }

\usepackage[dvips]{graphicx}
\usepackage{epsfig,amsmath,amssymb,verbatim,mathrsfs,array,layout,textcomp,amssymb,latexsym, slashed,graphicx}

\usepackage{graphicx}
\usepackage{amsmath}
\usepackage{amssymb}
\usepackage{verbatim,mathrsfs,array,layout,textcomp,latexsym, slashed}

\usepackage{amsmath, amssymb, graphicx, rotating}
\usepackage{hyperref}
\usepackage[usenames,dvipsnames]{color}

\newcommand{\beq}{\begin{eqnarray}}
\newcommand{\eeq}{\end{eqnarray}}

\def\beqa{\begin{eqnarray}}
\def\eeqa{\end{eqnarray}}

\newcommand{\bv}{\left(\begin{array}{c}}
\newcommand{\ev}{\end{array}\right)}
\newcommand{\bmtwo}{\left(\begin{array}{cc}}
\newcommand{\bmthree}{\left(\begin{array}{ccc}}
\newcommand{\emn}{\end{array}\right)}
\newcommand{\bmtwoc}{\left\{\begin{array}{cc}}
\newcommand{\bmthreec}{\left\{\begin{array}{ccc}}
\newcommand{\emnc}{\end{array}\right\}}
\newcommand{\ba}{\begin{array}}
\newcommand{\ea}{\end{array}}
\newcommand{\DM}{{\text{DM}}}

\newcommand{\TeV}{\text{ TeV}}
\newcommand{\MeV}{\text{ MeV}}

\newcommand{\half}{\frac{1}{2}}

\newcommand{\Teq}{T_{\text{eq}}}

\newcommand{\units}[1]{\mathrm{\; #1}}

\def\lsim{\mathrel{\rlap{\lower4pt\hbox{\hskip1pt$\sim$}}
     \raise1pt\hbox{$<$}}}         
\def\gsim{\mathrel{\rlap{\lower4pt\hbox{\hskip1pt$\sim$}}
     \raise1pt\hbox{$>$}}}         

\begin{document}

\vspace*{-30mm}
\font\mini=cmr10 at 0.8pt

\title{

The SIMP Miracle\\ \vspace{-.34cm}{\mini ====================================================================25\% of the authors prefer the title: `SIMP Dark Matter'.   They are uncomfortable with the term `miracle' in this scenario.  Damn democracy!==================================================================.}}

\author{Yonit Hochberg${}^{1,2}$}\email{yonit.hochberg@berkeley.edu}
\author{Eric Kuflik${}^3$}\email{ekuflik@post.tau.ac.il}
\author{Tomer Volansky${}^3$}\email{tomerv@post.tau.ac.il}
\author{Jay G. Wacker${}^4$}\email{jgwacker@stanford.edu}
\affiliation{${}^1$Ernest Orlando Lawrence Berkeley National Laboratory, University of California, Berkeley, CA 94720, USA}
\affiliation{${}^2$Department of Physics, University of California, Berkeley, CA 94720, USA}
\affiliation{${}^3$Department of Physics, Tel Aviv University, Tel Aviv, Israel}
\affiliation{${}^4$SLAC National Accelerator Laboratory, Stanford University, Menlo Park, CA 94025 USA}

\begin{abstract}We present a new paradigm for achieving thermal relic dark matter.  The mechanism arises when a nearly secluded dark sector  is thermalized with the Standard Model after reheating.  The freezeout process is a number-changing $3\rightarrow 2$ annihilation of strongly-interacting-massive-particles (SIMPs) in the dark sector, and points to  sub-GeV dark matter.  The couplings to the visible sector, necessary for maintaining thermal equilibrium with the Standard Model, imply measurable signals that will allow coverage of a significant part of the parameter space with future indirect- and direct-detection experiments and via direct production of dark matter at colliders.  Moreover, $3\to2$ annihilations typically predict sizable $2\to2$ self-interactions which naturally address the `core vs. cusp' and `too-big-to-fail' small structure problems.
\end{abstract}

\maketitle

\section{Introduction}
Dark matter (DM) makes up the majority of the mass in the Universe, however, its identity is unknown. The few properties known about DM are that it is cold and massive, it is not electrically charged, it is not colored and it is not very strongly self-interacting. One possibility for the identity of DM is that it is a thermal relic from the early Universe.
Cold thermal relics are predicted to have a mass
\begin{eqnarray}
m_\DM \sim   \alpha_{\rm ann} \left(T_{\rm eq} M_{\rm Pl} \right)^{1/2} \sim  \TeV\,,
\end{eqnarray}
where $\alpha_{\text{ann}}$ is the effective coupling constant of the $2\to 2$ DM annihilation cross section, taken to be of order weak processes $\alpha_{\rm ann}\simeq1/30$ above, $T_{\rm eq}$ is the matter-radiation equality temperature and $M_{\rm Pl}$ is the reduced Planck mass. The emergence of the weak scale from a geometric mean of two unrelated scales, frequently called the WIMP miracle, provides an alternate motivation beyond the hierarchy problem for TeV-scale new physics.

\begin{figure}[t!]
\begin{center}
\includegraphics[width=0.5\textwidth]{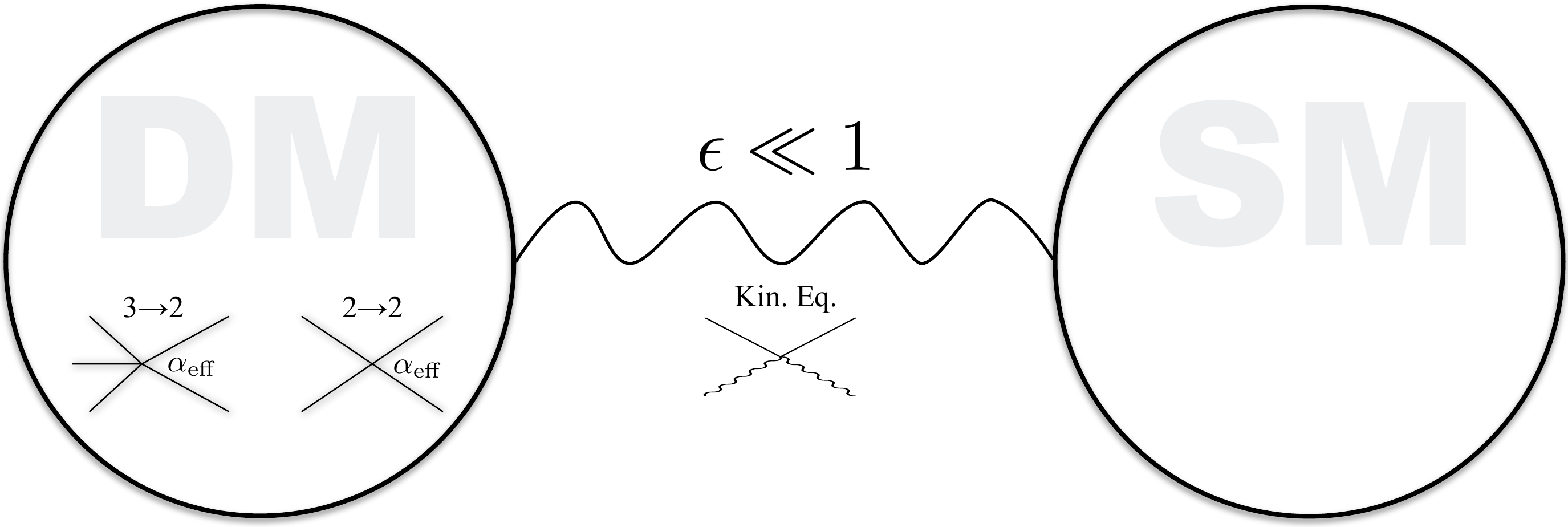}
\caption{
\label{Fig: scheme}
A schematic description of the SIMP paradigm.   The dark sector consists of DM which annihilates via a $3\to2$ process.   Small couplings to the visible sector allow for thermalization of the two sectors, thereby allowing heat to flow from the dark sector to the visible one.  DM self interactions are naturally predicted to explain small scale structure anomalies while the couplings to the visible sector predict measurable consequences.  }
\end{center}
\end{figure}

In this work we show that there is another mechanism that can produce thermal relic DM even if $\alpha_{\text{ann}} \simeq 0$. In this limit, while thermal DM cannot freeze out through the standard $2\to 2$ annihilation, it may do so via a $3\to 2$ process, where three DM particles collide and produce two DM particles. The mass scale that is indicated by this mechanism is given by a generalized geometric mean,
\begin{eqnarray}
m_\DM \sim \alpha_{\text{eff}} \left(T_{\rm eq}^2 M_{\rm Pl}\right)^{1/3} \sim 100 \units{MeV}\,,
\end{eqnarray}
 where $\alpha_{\text{eff}}$ is the effective strength of the self-interaction of the DM which we take as $\alpha_{\rm eff}\simeq 1$ in the above. As we will see, the $3\to2$ mechanism points to strongly self-interacting DM at or below the GeV scale. In similar fashion, a $4\to2$ annihilation mechanism, relevant if DM is charged under a $Z_2$ symmetry, leads to DM in the keV to MeV mass range. In this case, however, a more complicated production mechanism, such as freeze-out and decay, is typically needed to evade cosmological bounds.

If the dark sector does not have sufficient couplings to the visible sector for it to remain in thermal equilibrium, the $3\to2$ annihilations heat up the DM, significantly altering structure formation~\cite{Carlson:1992fn,deLaix:1995vi}. In contrast, a crucial aspect of the mechanism described here is that the dark sector is in thermal equilibrium with the Standard Model (SM), {\it i.e.} the DM has a phase-space distribution given by the temperature of the photon bath.  Thus, the scattering with the SM bath enables the DM to cool off as heat is being pumped in from the $3\to2$ process.
Consequently, the $3\to2$ thermal freeze-out mechanism generically requires measurable couplings between the DM and visible sectors.   A schematic description of the SIMP paradigm is presented in Fig.~\ref{Fig: scheme}.

The phenomenological consequences of this paradigm are two-fold.  First, the significant DM self-interactions have implications for structure formation, successfully addressing the so called `core vs. cusp' and `too big to fail' problems (see {\it e.g.}~\cite{Spergel:1999mh,deBlok:2009sp,BoylanKolchin:2011de}).
Second, the interactions between the DM and visible sectors predict significant direct and indirect signatures which may be probed in the near future.

In this letter, we aim to present a new paradigm for DM, rather than a specific DM candidate. For this reason, we do not explore particular models for the dark sector, but instead use a simplified effective description in order to understand the properties of the DM sector and its interaction with the SM such that the mechanism is viable. A detailed study exploring models for the SIMP mechanism is underway~\cite{simp2}.

\section{The $3\to 2$ Mechanism}\label{Sec: 3to2}
As mentioned above, the $3\to 2$ annihilation mechanism predicts a mass range for the DM, just as the standard $2\to 2$ annihilation mechanism  predicts the TeV scale. The estimate of the indicated mass scale is presented here, and is later verified by solving the Boltzmann equation explicitly.

It is useful to express quantities in the freeze-out estimate in terms of measured quantities.  In particular, the DM number density is given by
\begin{eqnarray}
  n_\DM = \frac{ \xi m_p \eta\, s}{m_\DM} = \frac{c\; \Teq\; s}{m_\DM} \,,
\end{eqnarray}
where $\xi = \rho_\DM/\rho_b \simeq 5.4$~\cite{Ade:2013zuv}, $m_p$ is the proton mass, $s$ is the entropy density of the Universe and $\eta$ is the baryon to entropy ratio. In the second equality, the number density is expressed in terms of the matter-radiation equality temperature, $\Teq\simeq 0.8\units{eV}$,
\begin{eqnarray}
\Teq = \frac{\xi}{c} m_p \eta\,, \qquad c\equiv \frac{\xi}{1+\xi}\frac{3}{4}\frac{g_{*,{\rm eq}}}{g_{*s,{\rm eq}}} \simeq ~0.54\,,
\end{eqnarray}
where $g_{*,{\rm eq}}$ and $g_{*s,{\rm eq}}$ are the energy and entropy effective number of relativistic  degrees of freedom at equality time.

Freeze out roughly occurs when the rate of the $3\to2$ process, $\Gamma_{3\to 2}$, is equal to the Hubble rate $H$. The $3\rightarrow 2$ rate is given by $\Phi^2 \sigma_{3\to 2}$, where $\Phi$ is the flux of particles incident on a particle and $\sigma_{3\to 2}$ is the `cross section' for the $3\to 2$ process. Using this, together with $\Phi=n v$, with $v$ parameterizing the average relative velocity between the colliding particles, the freeze-out condition is given by
\begin{eqnarray}
n_\DM^2 \langle  \sigma_{3\rightarrow 2} v^2 \rangle|_{T=T_F} =0.33 \sqrt{g_{*,F}}\frac{T^2_F}{M_{\rm Pl}}\, .
\end{eqnarray}
We parameterize the $3\rightarrow 2$ cross section by
\begin{eqnarray}
\label{eq:alphaeff}
 \langle\sigma v^2\rangle_{3\rightarrow 2} \equiv \frac{ \alpha_{\rm eff}^3 }{m_{\text{DM}}^5}\,,
\end{eqnarray}
where $\alpha_{\rm eff}$ is the effective coupling strength entering the {\em thermally averaged} cross section. We stress that the effective coupling above can be significantly larger than unity if, for example, the number of DM degrees of freedom is large, if the cross-section is non-perturbatively enhanced, or if the $3\rightarrow 2$ process is mediated by a light particle.

The rest of the freeze-out estimate proceeds in a straight forward manner. Using (see {\it e.g.}~\cite{Kolb:1990vq})
\begin{eqnarray}
s = \frac{\kappa}{c} T^3 \qquad \kappa = \frac{2\,\pi^2\,c\,g_{*s}(T)}{ 45}\,,
\end{eqnarray}
and parameterizing the freeze-out temperature as
\begin{eqnarray}
T_F  =\frac{m_\DM}{x_F}\,,
\end{eqnarray}
the DM mass indicated by the $3\rightarrow 2$ process is
\begin{eqnarray}
\label{Eq: mass}
m_\DM \simeq 1.4 \,\alpha_{\rm eff} x_F^{-1} \left(  g_{*,F}^{-\half}\;x_F^{-1}\;  ( \kappa \; \Teq)^2  M_{\rm Pl} \right)^{\frac{1}{3}}\, .
\end{eqnarray}
Taking  $x_F = 20$ and $\alpha_{\rm eff}=1$ for a (rather) strongly interacting theory that freezes out while the DM is non-relativistic, we arrive at
\begin{eqnarray}
m_\DM \simeq 40 \MeV \qquad  \mathrm{(3\to 2)}\,.
\end{eqnarray}
Small corrections are found when the more precise Boltzmann equations are solved (see Fig.~\ref{Fig: mass}). Thus in analogy to the standard thermal WIMP, where weak coupling gives rise to the weak scale, the $3\rightarrow 2$ freezeout mechanism gives rise to strong-scale DM for strong coupling.  Lower (higher) DM mass is of course consistent with lower (higher)  $\alpha_{\rm eff}$.  As we will see,
 self-interactions of DM along with CMB and BBN constraints point to the strongly interacting limit of large $\alpha_{\rm eff}$.   We thus dub this scenario the {\em Strongly Interacting Massive Particle} (SIMP) paradigm (to be distinguished from DM models with strong interactions with the visible sector, commonly called SIDM~\cite{Goldberg:1986nk,Rich:1987st,Chivukula:1989cc,Starkman:1990nj,Nardi:1990ku,Mohapatra:1997sc,Mohapatra:1999gg,Teplitz:2000zd}).

If DM is a fermion or if the   dark sector  admits   a $Z_2$ symmetry, the   $3\to2$ annihilation process is forbidden.    Consequently, freeze-out can proceed via a $4\to2$ annihilation channel. For such a case with a cross section that is parameterized as
\begin{eqnarray}
\langle\sigma v^3\rangle_{4\rightarrow 2} \equiv \frac{ \alpha_{\rm eff}^4}{ m_\DM^8}\,,
\end{eqnarray}
a similar estimate results in a DM mass of
\begin{eqnarray}
\label{Eq: mass2}
m_\DM \simeq 1.3 \,\alpha_{\rm eff} x_F^{-2} \left(  g_{*,F}^{-\half}\; x_F \;(\kappa\; \Teq)^3  M_{\rm Pl}\right)^{\frac{1}{4}}\,,
\end{eqnarray}
which, for $x_F = 14$ (as obtained by solving the Boltzmann equation)   and $\alpha_{\rm eff} = 1$, points to DM mass of order
\begin{eqnarray}\label{Eq: mass2num}
m_\DM \simeq 100 \units{keV} \qquad  \mathrm{(4\to 2)}\,.
\end{eqnarray}
Such light DM in thermal equilibrium with electrons, photons or neutrinos is mostly excluded by BBN and CMB data, and is in tension with structure formation (see below).
  It is possible, however, to evade the latter bounds if, for example, the frozen out specie above is not the DM and instead decays to a lighter stable state, much as in the case of a superWIMP~\cite{oai:arXiv.org:hep-ph/0302215}. Finally, higher $n\to 2$ freeze-out interactions yield sub-keV DM masses and are omitted from further discussion.

\section{Thermal Equilibrium}
\label{Sec: Thermal}
Throughout the above estimate, we have assumed that the dark sector and SM remained in thermal equilibrium. However, the processes that keep the two sectors in thermal equilibrium are the crossing diagrams of the processes that lead to  $2\to2$ annihilation into the SM.  Thus, the assumption of thermal equilibrium might naively imply that the dominant number-changing process for the DM is the $2\to2$ annihilation channel.  In this section, we find the condition under which the latter is subdominant while thermal equilibrium is maintained. These will be the conditions under which the $3\to2$ mechanism is viable.

The ratio of the scattering rate off of SM particles $\Gamma_{\text{kin}}$ and the annihilation rate to SM particles $\Gamma_{\text{ann}}$ is
\begin{eqnarray}
\frac{\Gamma_{\text{kin}} }{\Gamma_{\text{ann}} }=\frac{n_{\rm SM} \langle  \sigma v\rangle_{\rm kin}}{n_{\rm DM} \langle  \sigma v\rangle_{\rm ann}}\simeq \frac{g_{{\rm SM},F}\, m_{\rm DM}}{\pi^2 \kappa\, \Teq } \simeq 5\times 10^{6}
\end{eqnarray}
where $g_{{\rm SM},F}$ is the effective number of relativistic SM degrees of freedom participating in the $2\to2$ annihilation process at freeze-out, the second equality uses $\langle  \sigma v\rangle_{\rm kin} \sim \langle  \sigma v\rangle_{\rm ann}$, and the last equality is derived for $m_{\rm DM} = 40\units{MeV}$. This large ratio is simply  understood by the sub-dominance of the DM number density at $T_F\gg T_{\rm eq}$.   Thus, if the SM couples to the DM, the process keeping these two sectors in kinetic equilibrium does not have to be changing the annihilation rate. A similar statement holds in the  standard thermal WIMP scenario.

  In order for the $3\to 2$ process to control freeze-out, while  not heating up the DM, the following inequalities must be satisfied up until freeze-out occurs:
\beq
\left. \frac{\Gamma_{\text{kin}}}{\Gamma_{3\rightarrow 2}} \right|_{T=T_F} &\gtrsim& 1,\label{Eq:cond1}\\
\left. \frac{\Gamma_{\text{ann}}}{\Gamma_{3\rightarrow 2}} \right|_{T=T_F}&\lesssim& 1\label{Eq:cond2}.
\eeq
We  parameterize the DM-SM scattering by a small coupling, $\epsilon$, with the relevant energy scale $m_{\rm DM}$, such that
\begin{eqnarray}
\langle  \sigma v\rangle_{\rm kin} \sim \langle  \sigma v\rangle_{\rm ann}\equiv \frac{\epsilon^2}{m_\DM^2}\,.
 \label{Eq:epsilondef}
\end{eqnarray}
The exact relation between the above cross sections can be calculated for particular couplings to SM particles. The first inequality, Eq.~\eqref{Eq:cond1}, which ensures that the coupling to the SM is strong enough to keep the dark sector and visible sector at a unified temperature, requires
\begin{eqnarray}\label{Eq: emin}
 \epsilon  \gtrsim  \epsilon_{\rm min} \equiv 2\ \alpha_{\rm eff}^{1/2}  \left(\frac{T_{\rm eq}}{M_{\rm Pl}}\right)^{1/3}  \simeq  1\times 10^{-9} \,,
\end{eqnarray}
where  the numerical estimates use  $\alpha_{\rm eff} = 1$ and  $x_F\simeq 20$, as will be justified by solving the Boltzmann equation explicitly. The second condition, ensuring that the annihilation of the dark sector to SM states is not efficient at freeze-out, implies
\begin{eqnarray}
\label{Eq: emax}
\epsilon \lesssim   \epsilon_{\rm max} \equiv 0.1\ \alpha_{\rm eff} \left(\frac{T_{\rm eq}}{M_{\rm Pl}}\right)^{1/6} \simeq 3 \times10^{-6}\,,
\end{eqnarray}
with the same choice of parameters as above. We learn that there is a large range of couplings of the DM to the SM in which Eqs.~\eqref{Eq:cond1} and \eqref{Eq:cond2} are satisfied.

\section{Solving the Boltzmann Equation}
\label{Sec: Numerics}
\begin{figure}[t!]
\begin{center}
\includegraphics[width=0.5\textwidth]{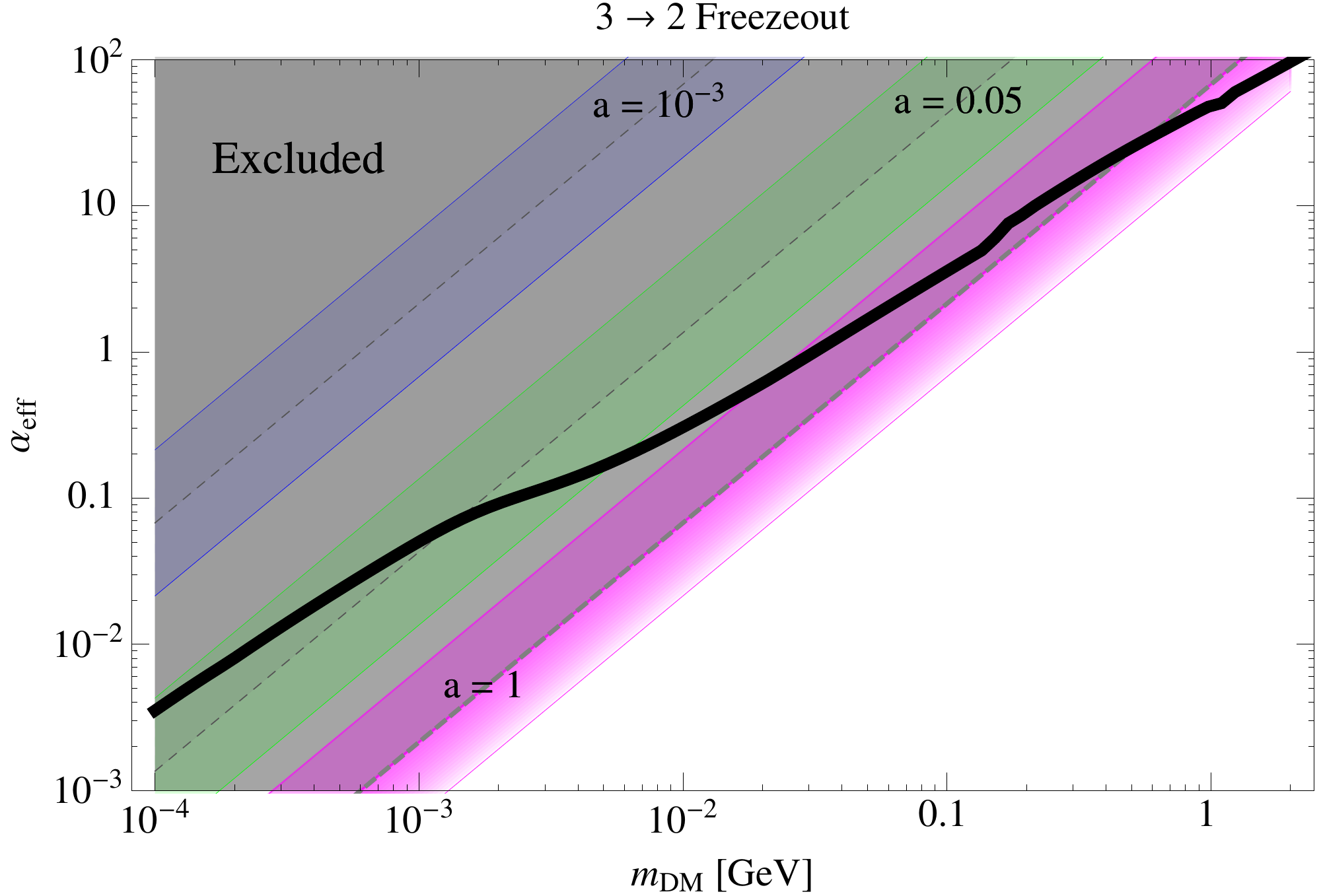}
\caption{
\label{Fig: mass}
$\alpha_{\rm eff}$ as a function of the DM mass (black solid line), derived from the numerical solution to the Boltzmann Equation in the $3\to2$ freeze-out scenario.   The colored regions show the preferred region as hinted by the `core vs. cusp' and `too-big-to-fail' small-scale structure anomalies for $a =1$ (magenta), $a=0.05$ (green), and $a=10^{-3}$ (blue). The region above the gray-dashed lines is excluded by the bullet-cluster~\cite{Clowe:2003tk,Markevitch:2003at,Randall:2007ph} and halo shape~\cite{Rocha:2012jg,Peter:2012jh} constraints, for each value of $a$. The shaded  gray  shows the exclusion region for $a=1$.}
\end{center}
\end{figure}
Thus far we have described the general setup of the SIMP mechanism, and we now move on to validating our results. There are a variety of scattering processes that are relevant to the Boltzmann equation of the $3\to 2$ system. In the following, $\chi$ represents the DM.

First, there is a self-interaction process of the $2\to2$ scattering within the dark sector, which does not change the number density of the DM. It does, however, ensure that the DM all remain at one single temperature. Second, there is the $3\rightarrow 2$
 number-changing process responsible for freeze-out, which sets the chemical potential of $\chi$ to zero. Assuming that the DM is strongly coupled, these two processes are fast, meaning that the DM number density $f(t,E)$ per phase space at a given time follows
\begin{eqnarray}
f(E,t) \propto n(t) e^{ -E/T_{\text{DM}}(t)}\,,
\end{eqnarray}
where $T_{\text{DM}}$ is the time-dependent temperature of the system which may differ from  the temperature of the SM. The third process is the scattering of DM off of the SM, which does not change the DM number density. If this process is active, then the SM and dark sector will have the same temperature, $T_{\rm DM}=T_{\rm SM}$. This kinetic equilibrium is crucial for ensuring that the dark sector does not stay hot. Finally, there is the standard annihilation process into SM particles, related by crossing symmetry to the above SM scattering process. The rate of this process must be subdominant to the $3\to2$ process at freeze-out, otherwise the standard DM freeze-out paradigm controls the relic abundance. Ensuring that this is small enough while maintaining the kinetic equilibrium rate large enough is a critical step in making the $3\rightarrow 2$ mechanism viable.

Combining the above, we arrive at the Boltzmann equation for the number density $n$ of $\chi$,
\begin{eqnarray}\label{Eq: boln}
\nonumber
\partial_t n + 3 H n &=& - \left(n^3 - n^2 n_{\text{eq}} \right)\langle \sigma v^2 \rangle_{3\to2}\\
&&
\quad -  \left( n^2 - n_{\text{eq}}^2 \right) \langle \sigma v \rangle_{\rm ann}\,.
\end{eqnarray}
Using the parametrization of Eqs.~\eqref{eq:alphaeff} and~\eqref{Eq:epsilondef}, and numerically integrating Eq.~\eqref{Eq: boln}, results in the nearly linear relationship between the 3-point self-coupling strength, $\alpha_{\rm eff}$, and the mass of the DM, $m_{\rm DM}$, shown by the black solid curve   in Fig.~\ref{Fig: mass}. We find $x_F \sim 14-24$  for the entire mass range of interest, and the results agree very well with the estimate of Eq.~\eqref{Eq: mass}.  We further find that compared to standard freeze-out from a $2\rightarrow2$ annihilation process, the SIMP begins freeze-out slightly later, but reaches the final relic abundance faster. The reason is that the back-reaction of the $2\rightarrow3$ process quickly becomes negligible while the $3\rightarrow2$ rate is proportional to $n^2$, rather than to $n$.

The cautious reader might note the absence of $\epsilon$ in the above results. The reason is that as long as Eqs.~\eqref{Eq:cond1} and~\eqref{Eq:cond2} are satisfied, the precise value of $\epsilon$ has little effect on the solutions $(\alpha_{\rm eff}, x_F, m_{\rm DM})$ to the Boltzmann equation. Given particular couplings between the dark sector and the SM, the range for $\epsilon$ can be found such that Eqs.~\eqref{Eq:cond1} and~\eqref{Eq:cond2} are indeed satisfied.

\section{A Toy Model}\label{Sec: toy}
To better understand the SIMP paradigm, we now present a weakly-coupled toy model for the dark sector which incorporates the $3\to2$ mechanism and leads to stable DM. Consider a $Z_3$-symmetric theory with a single scalar, $\chi$, defined by
\beq
\label{eq:toy}
{\cal L}_{\rm DM} =|\partial \chi|^2-m_{\rm DM}^2 |\chi|^2-\frac{\kappa}{6}\chi^3-\frac{\kappa^\dagger}{6}\chi^{\dagger3}-\frac{\lambda}{4}|\chi|^4\,.
\eeq
With the above couplings, tree-level $2\to2$ self-interactions and $3\to2$ scattering are induced. For a single scale model, defining $g$ via $\kappa=g\, m_{\rm DM}$ and taking $\lambda\sim g^2$, the $2\to2$ scattering cross section scales as $g^4/m_{\rm DM}^2$, and the $3\to2$ one as $g^6/m_{\rm DM}^5$, motivating our parametrization of Eq.~\eqref{eq:alphaeff}. The stability of the DM is guaranteed by the global symmetry.

Let us now introduce small interactions between the DM and the visible sector.  As an example, consider first an interaction with SM fermions $f$,
\begin{equation}
\label{eq:Lint}
{\cal L}_{\rm int} = \frac{m_f}{\Lambda^2}\chi^\dagger \chi \overline f f\,,
\end{equation}
which induces both $2\to2$ annihilations and scatterings.  Identifying the $\epsilon$ defined in Eq.~\eqref{Eq:epsilondef} to be of order $\epsilon\simeq {\cal O}(m_f m_{\rm DM}/\Lambda^2)$, the $2\to2$ annihilation rate is negligible while kinetic equilibrium is maintained, for $\epsilon_{\rm min} \lesssim \epsilon  \lesssim \epsilon_{\rm max}$.   One may further check that annihilations such as  $\chi\chi f\to \chi^\dagger f$, which are induced by the interactions in Eqs.~\eqref{eq:toy} and~\eqref{eq:Lint}, are also negligible despite the large number density in the thermal bath.
Alternatively, the dark sector may couple to the visible one through photons,
\begin{equation}
\label{eq:Lint2}
{\cal L}_{\rm int} = \frac{\alpha_{\rm EM}}{4\pi \Lambda^2}\chi^\dagger \chi F_{\mu\nu}F^{\mu\nu}\,,
\end{equation}
in which case, $\epsilon \simeq {\cal O}(\alpha_{\rm EM} m_\DM^2/4\pi \Lambda^2)$.

To conclude, we find that for a low-scale (of order, say, 100~MeV) dark sector with DM  described by Eq.~\eqref{eq:toy}, the correct relic abundance is obtained if the sector communicates with the visible one (say, through couplings to electrons, muons or photons) via a new scale in the GeV to 10's of TeV range.  Such mediators are thus constrained by LEP~\cite{Fox:2011fx,Essig:2013vha}  (see Fig.~\ref{fig:epsbounds}) and are expected to be within reach of ongoing collider experiments.  As we now discuss, such a sector will not only have experimental signatures that will allow discovery, but will also have unavoidable self-interactions which can address long-standing puzzles in structure formation.

\section{Signatures}
\label{Sec: Signatures}
\begin{figure*}[ht!]
\includegraphics[width=1\textwidth]{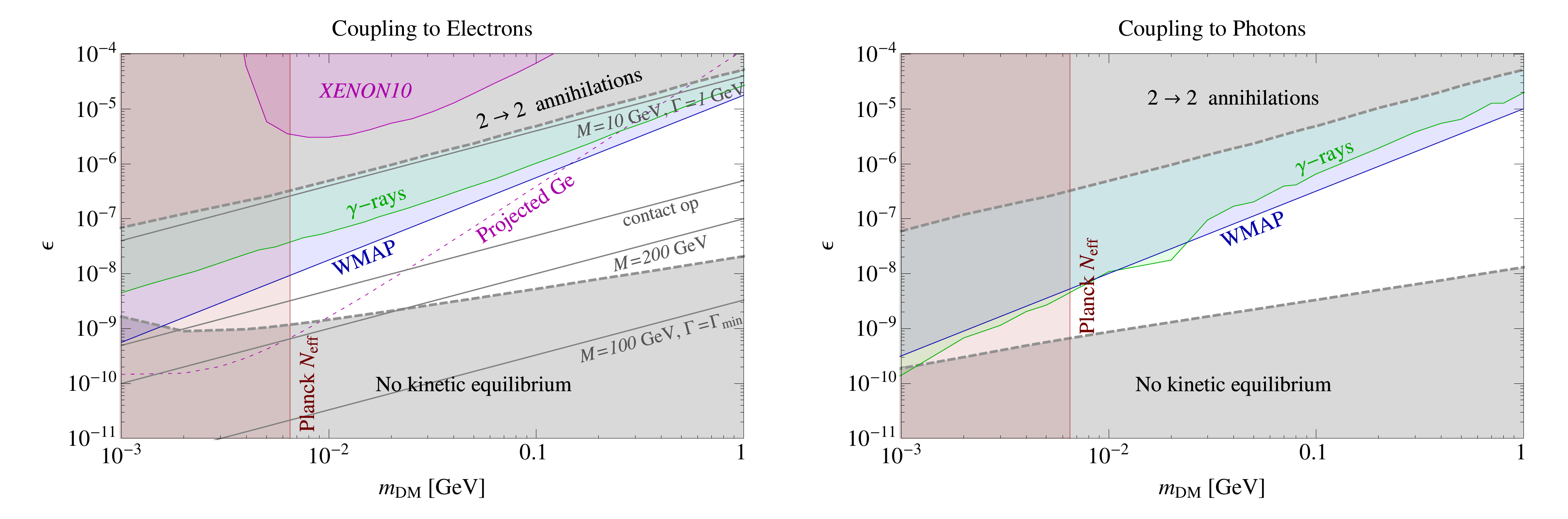}
\caption{\label{fig:epsbounds}
The bounds on $\epsilon$ vs. $m_{\rm DM}$. {\bf Left, coupling to electrons:} The grey regions (outlined by thick dashed lines) represents the range of parameters in which kinetic equilibrium with the SM is not maintained (lower gray region), and where the standard $2\to2$ annihilation to the SM is not subdominant to the $3\to 2$ process (upper gray region). Also shown are the exclusion limits from: direct-detection in Xenon10~\cite{Essig:2012yx} (purple region), along with the expected future bound from a germanium-based electron recoil experiment~\cite{Essig:2011nj} (dashed-purple); CMB and low red shift data constraints for electrons~\cite{Madhavacheril:2013cna} (blue region); modification of $N_{\rm eff}$~\cite{Boehm:2013jpa} (red region); indirect detection of $\gamma$-rays~\cite{Essig:2013goa} (green region); direct production at LEP for a variety of mediator mass, $M$, and width, $\Gamma$ (solid-gray)~\cite{Fox:2011fx}.
{\bf Right, coupling to photons:} The grey regions (outlined by thick dashed lines) represents the range of parameters in which kinetic equilibrium with the SM is not maintained (lower gray region), and where the standard $2\to2$ annihilation with the SM is not subdominant to the $3\to 2$ process (upper gray region). Also shown are the exclusion limits from: indirect detection of $\gamma$-rays~\cite{Essig:2013goa} (green region); conservative CMB and low red shift data constraints~\cite{Madhavacheril:2013cna} (blue region); modification of $N_{\rm eff}$~\cite{Boehm:2013jpa} (red region).
}
\end{figure*}

The paradigm discussed in this letter not only provides a new mechanism for producing the DM relic abundance, but also predicts interesting and measurable signatures. There are two distinct reasons for this. First, much as in the standard thermal WIMP scenario, the DM must be in thermal equilibrium with the visible sector. Consequently, it must have non-negligible couplings to SM particles, which in turn  predict observable signals.  Second, the non-vanishing 5-point interaction required for the $3\to2$ annihilations also implies sizeable self-couplings which alter the predictions for structure formation. Below, we briefly summarize these two aspects, postponing many of the details to future work~\cite{simp2}.

We begin with structure formation. The persistent failure of N-body simulation to reproduce the small-scale structure of observed galactic halos has led to the `core versus cusp' and `too big to fail' problems. This motivates self-interacting DM with a strength~\cite{ Zavala:2012us,Vogelsberger:2012ku,Rocha:2012jg,Peter:2012jh}
\beq\label{eq:strform}
\left(
\frac{\sigma_{\rm scatter}}{m_{\rm DM}}
\right)_{\rm obs}
=(0.1-10) \units{cm^2/g}  \,.
\eeq
On the other hand, bullet-cluster constraints~\cite{Clowe:2003tk,Markevitch:2003at,Randall:2007ph} as well as recent simulations which reanalyze the constraints  from halo shapes~\cite{Rocha:2012jg,Peter:2012jh}, suggest the limits on the DM self-interacting cross section (at velocities $\gtrsim300\units{km/sec}$)   are
\begin{equation}
 \label{eq:bullet}
\frac{\sigma_{\rm scatter}}{m_{\rm DM}} \lesssim 1 \units{cm^2/g}\,.
\end{equation}
The above constraint leaves a viable region for the preferred strength of DM self-interactions.

The SIMP scenario naturally predicts a sizable contribution to the above $2\rightarrow 2$ scatterings.  One may parametrize it by defining $a \equiv \alpha_{2\to2} / \alpha_{\rm eff}$, such that
\beq\label{Eq:scat}
\frac{\sigma_{\rm scatter}}{m_{\rm DM}} = \frac{a^2 \alpha_{\rm eff}^2}{m_{\rm DM}^3}\,,
\eeq
and one expects $a$ to be of order unity.
This can be readily checked for the toy model discussed above, where $a={\cal O}(1)$ is found for a wide range of values of the couplings of~Eq.~\eqref{eq:toy}. For the $3\to2$ SIMP scenario, the constraint, Eq.~\eqref{eq:bullet}, points to the strongly interacting regime with DM masses at or below the GeV scale. Interestingly, this region in parameter space automatically solves the  small-structure anomalies discussed above.  Indeed, one may  use Eqs.~\eqref{eq:strform} and~\eqref{eq:bullet} together with the relation Eq.~\eqref{Eq: mass} to derive a preferred range of $\alpha_{\rm eff}$. Taking into account the numerical corrections as found using the Boltzmann equation, we arrive at
\begin{equation}
\label{eq:alphaSelf}
0.3 \left(\frac{a}{0.2}\right)^2 \lesssim \alpha_{\rm eff} \lesssim 8 \left(\frac{a}{0.2}\right)^2\,,
\end{equation}
where the lower bound above arises from the upper bound of Eq.~\eqref{eq:bullet}.   The corresponding DM mass is in the range of $8\left(\frac{a}{0.2}\right)^2\units{MeV} \lesssim m_{\rm DM} \lesssim 200\left(\frac{a}{0.2}\right)^2\units{MeV}$.
 In Fig. \ref{Fig: mass} we show the full region preferred by the small-scale structure anomalies, and the region excluded by bullet-cluster and halo-shape constraints. The colored regions show the preferred region for $a =1,~0.05,~10^{-3}$. The region above the corresponding  gray-dashed lines is excluded by the bullet-cluster and halo shape constraints, for each value of $a$. The solid gray region shows that  exclusion region for $a=1$.

Models of strongly interacting DM that can accommodate the structure formation anomalies have been proposed in the literature~\cite{Kribs:2009fy,Alves:2009nf,Cline:2013zca,Kumar:2011iy,Alves:2010dd, Lisanti:2009am, CyrRacine:2012fz, Kaplan:2009de,Tulin:2013teo,Kaplinghat:2013kqa,Boddy:2014yra}, however most of them rely upon either a new long range force or a non-thermal mechanism to explain the DM relic abundance (see~\cite{Vogelsberger:2012sa,Kahlhoefer:2013dca,Feng:2009hw,Fan:2013yva,Cline:2013pca} for additional constraints that arise with long range forces). In contrast, the SIMP mechanism offers simplicity in the generation of the relic density and naturally points to the correct scale of self-interactions once the relic abundance is fixed to the observed value.

We now move on to the constraints on the coupling between the SIMP and SM particles. In addition to those of Eqs.~\eqref{Eq:cond1} and~\eqref{Eq:cond2}, there are constraints from direct detection, indirect detection and cosmological data. To this end, we consider separately effective couplings of the SIMP to electrons or photons:
\begin{enumerate}
  \item {\bf Coupling to electrons}. We take the interaction of Eq.~\eqref{eq:Lint} with $f=e^-$. Bounds on $\epsilon$, defined through Eq.~\eqref{Eq:epsilondef}, as a function of the mass, are then derived from (I)~the requirements of Eqs.~~\eqref{Eq:cond1} and~\eqref{Eq:cond2}; (II)~Xenon10 electron ionizations data~\cite{Essig:2012yx} and the projection for a germanium-based electron recoil experiment~\cite{Essig:2011nj}; (III)~CMB data~\cite{Madhavacheril:2013cna}; (IV)~modification to neutrino $N_{\rm eff}$~\cite{Boehm:2013jpa} from Planck data \cite{Ade:2013zuv}; (V)~indirect detection of FSR radiation off the $\chi\chi\to ee$ process~\cite{Essig:2013goa}; and (VI)~direct production constraints from LEP~\cite{Fox:2011fx}. Our results are depicted in the left panel of Fig.~\ref{fig:epsbounds}. Constraints from supernovae cooling~\cite{Raffelt:1996wa,Dreiner:2003wh,Dreiner:2013mua} are not depicted as they are irrelevant in the allowed parameter space.

  \item {\bf Coupling to photons}. We now take the  interaction of Eq.~\eqref{eq:Lint2}.   The relevant bounds on $\epsilon$ in this case come from (I)~Eqs.~~\eqref{Eq:cond1} and~\eqref{Eq:cond2}; (II)~indirect detection of annihilation into photons~\cite{Essig:2013goa}; (III)~CMB data~\cite{Madhavacheril:2013cna} (assuming an unsuppressed absorption efficiency~\cite{Padmanabhan:2005es}); and (IV)~modification to effective number of degrees of freedom $N_{\rm eff}$~\cite{Boehm:2013jpa} from Planck data \cite{Ade:2013zuv}.  Our results are depicted in the right panel of Fig.~\ref{fig:epsbounds}. Constraints from electron ionization data at Xenon10 are not depicted since they arise either from a photon loop, in which case the bound is weak, or from a tree level process to an $e^+ e^- \gamma$ final state, in which case a dedicated study is required due to the dependence of the form factor on the momenta of the outgoing photon.
\end{enumerate}
We comment that the Planck $N_{\rm eff}$ bound can be evaded if the DM couples simultaneously to electrons, photons and neutrinos, in which case a lower bound on the DM mass of ${\cal O}({\rm MeV})$ arises from BBN, unless the DM is a real scalar~\cite{Boehm:2013jpa}. Even in this case, note that the $4\to2$ mechanism described by Eq.~\eqref{Eq: mass2} is in tension with structure formation constraints unless more complicated scenarios, such as freeze-out and decay, are invoked.

\section{Conclusions}
\label{Sec: Conclusions}
In this work we proposed a new paradigm for thermal dark matter. The requisite features of the mechanism are the following:
\begin{itemize}
  \item Freeze-out proceeds through a self-interacting $3\to2$ annihilation within the dark sector.
  \item At freeze-out, the dark sector can exchange heat with the SM plasma.
  \item The number-changing standard annihilation of two DM particles into two SM particles is negligible compared to the $3\to2$ process at freeze-out.
\end{itemize}
Under this setup, the mechanism robustly predicts dark matter at or below the GeV scale with strong self interactions.  The exact relation between the mass and coupling of the DM system are depicted in Fig.~\ref{Fig: mass}, with freeze-out temperature $T_F\sim m_{\rm DM}/20$ throughout the mass range of interest.

The primary challenge in realizing this mechanism is in guaranteeing that the dark sector stays in kinetic equilibrium with the SM sector while the standard annihilation process is suppressed.  There is, however, no intrinsic obstacle to achieving this. The conditions under which the setup is viable are given in Eqs.~\eqref{Eq: emin} and ~\eqref{Eq: emax}. The self interaction of this SIMP mechanism unavoidably contributes to the cross section relevant for structure formation in a sizable way. Moreover, the couplings between the dark and visible sectors imply, much as in the WIMP scenario, measurable experimental signatures that will allow the discovery of such DM with upcoming experiments.  Constraints on the effective coupling between the DM and the SM as a function of the DM mass arising from direct detection, indirect detection, cosmology and direct production at colliders, are depicted in Fig.~\ref{fig:epsbounds} for coupling to electrons (left) and photons (right).

Significant work remains to study this intriguing scenario in detail.   More concrete realizations as well as a detailed study of the experimental constraints will be presented in an upcoming publication~\cite{simp2}.

\mysections{Acknowledgments}
We thank Nima Arkani-Hamed, Lawrence Hall, Maxim Perelstein, Tracy Slatyer and Jesse Thaler for useful discussions. We especially thank Jeremy Mardon for useful discussions and comments on the manuscript.  JW thanks Tomas Rube for initial collaboration on this project in 2008.   The work of YH is supported in part by the U.S. National Science Foundation under Grant No. PHY-1002399. YH is an Awardee of the Weizmann Institute of Science - National Postdoctoral Award Program for Advancing Women in Science. EK and TV are supported in part by a grant from the Israel Science Foundation. TV is further supported by the US-Israel Binational Science Foundation, the EU-FP7 Marie Curie, CIG fellowship and by the I-CORE Program of the Planning Budgeting Committee and the Israel Science Foundation (grant NO 1937/12). JW is supported by the DOE under contract number DE-AC02-76-SF00515.


\end{document}